\documentstyle[prl,aps,amssymb,amsmath,epsf,multicol,psfrag]{revtex}

\begin{document}

\begin{multicols}{2}

\noindent
{\bf Comment on `Experimental Entanglement Swapping: Entangling Photons That 
	Never Interacted' \cite{pieter}}

\medskip

\noindent
Entanglement swapping has been experimentally demonstrated by Pan {\em et 
al}.\ using two parametric down-converters and linear optical elements 
\cite{pan1998}. This experiment was first proposed (in a slightly modified 
form) by Zukowski, Zeilinger, Horne and Ekert \cite{zzhe1993}, who stressed 
the possibility of so-called {\em event-ready} detections using entanglement
swapping. Indeed, Pan {\em et al}.\ note that their experiment for the first 
time gives the possibility of event-ready detections. However, in this comment 
we will show that with the current state of technology event-ready detections 
can not be performed this way.

For event-ready detections we need a way of deciding that a maximally 
polarisation entangled state left the apparatus. This is equivalent to having 
an outgoing state 
\begin{equation}\label{ere}
 \rho_{\rm out} = |{\rm Bell}\rangle\langle{\rm Bell}| + O(\xi) 
\end{equation}
conditioned on detector coincidences, with $|{\rm Bell}\rangle$ any of 
the four polarisation Bell states ($|\Psi^{\pm}\rangle$, $|\Phi^{\pm}\rangle$) 
and $\xi \ll 1$.

In Refs.\ \cite{pan1998,zzhe1993}, entanglement swapping is described in terms 
of two anti-symmetric polarisation Bell states $|\Psi^-\rangle_{ab}\otimes
|\Psi^-\rangle_{cd}$ which, after a Bell detection of modes $b$ and $c$ is 
turned into a Bell state in modes $a$ and $d$. In the case of the experiment 
of Pan {\em et al}.\ (see Fig.\ 1) modes $b$ and $c$ are sent into a 
beam-splitter. A coincidence in the detectors $D_u$ and $D_v$ identifies a 
$|\Psi^-\rangle$ Bell state. Modes $a$ and $d$ should now be in the $|\Psi^-
\rangle$ Bell state as well. However, as has been pointed out previously 
\cite{pan1998,zzhe1993,braun1998,kok1999}, parametric down-converters do {\em 
not} produce $|\Psi^-\rangle$ Bell states. Instead, there is a strong 
pollution of vacuum and small contributions from higher down-conversions which 
invalidate the above description of entanglement swapping. 

To lowest non-trivial order the (unnormalised) states (with linear 
polarisations along $x$ and $y$ axes) leaving the apparatus after a two-fold 
coincidence conditioned on the four polarisation settings in $D_u$ and $D_v$ 
are:
\begin{mathletters}\label{sep1}
\begin{eqnarray}
 |\phi_{(x,x)}\rangle_{\rm ad} &=& |0,y^2\rangle - |y^2,0\rangle \\
 |\phi_{(x,y)}\rangle_{\rm ad} &=& |0,xy\rangle - |y,x\rangle + |x,y\rangle -
   |xy,0\rangle \\
 |\phi_{(y,x)}\rangle_{\rm ad} &=& |0,xy\rangle + |y,x\rangle - |x,y\rangle -
   |xy,0\rangle \\
 |\phi_{(y,y)}\rangle_{\rm ad} &=& |0,x^2\rangle - |x^2,0\rangle\; ,
\end{eqnarray}
\end{mathletters}
where $|\phi_{(i,j)}\rangle_{\rm ad}$ is the outgoing state conditioned on an
$i$-polarised photon in $D_u$ and a $j$-polarised photon in $D_v$ ($i,j \in \{ 
x,y \}$). Here, for instance, $|y^2\rangle$ is a $y$-polarised mode in
a 2 photon Fock state. No polarisers were used in the Bell state detection of 
Pan {\em et al}., and the state leaving the apparatus is a random mixture 
$\rho$ of these four states. This mixed state is different from the outgoing 
state Pan {\em et al}.\ describe in Ref.\ \cite{pan1998}. 

Using the Peres-Horodecki partial transpose criterion \cite{peres} it can be 
shown that $\rho$ is indeed entangled ($\rho$ has negative eigenvalues). 
However, it can not be used for event-ready detections of polarisation 
entanglement since the states in Eq.\ (\ref{sep1}) are not of the form of Eq.\ 
(\ref{ere}). 

Can we turn any of the states in Eq.\ (\ref{sep1}) into the form of Eq.\
(\ref{ere})? Additional photon sources are not allowed since that would take 
us beyond the entanglement swapping protocol. It can 
easily be verified that there is no linear optical transformation which takes 
any of the states in Eq.\ (\ref{sep1}) to any of the four Bell states. Still 
we need to bring a two-photon state to a two-photon Bell state. Ordinary 
detectors destroy photons, so we need at least sufficiently good {\em 
polarisation independent} quantum non-demolition ({\sc qnd}) measurements or a
quantum computer of some kind. Furthermore, the correlations (constituting the 
entanglement) must be preserved. Such {\sc qnd} detectors correspond to 
technology not yet available. This means that event-ready detections are not 
yet possible using entanglement swapping with parametric down-conversion.

To conclude, we have calculated the state leaving the apparatus of the 
entanglement swapping experiment by Pan {\em et al}. Contrary to the
description of the outgoing state in Ref.\ \cite{pan1998}, this is a random 
mixture of four (entangled) states. None of them is suitable for event-ready 
detections with present technology.

This research is funded in part by EPSRC grant GR/L91344.

\begin{figure}
 \begin{center}
  \begin{psfrags}
     \psfrag{a}{$a$}
     \psfrag{b}{$b$}
     \psfrag{c}{$c$}
     \psfrag{d}{$d$}
     \psfrag{BS}{\small BS}
     \psfrag{DC 1}{\small DC1}
     \psfrag{DC 2}{\small DC2}
     \psfrag{u}{$D_u$}
     \psfrag{v}{$D_v$}
     \epsfxsize=8in
     \epsfbox[-100 35 1000 150]{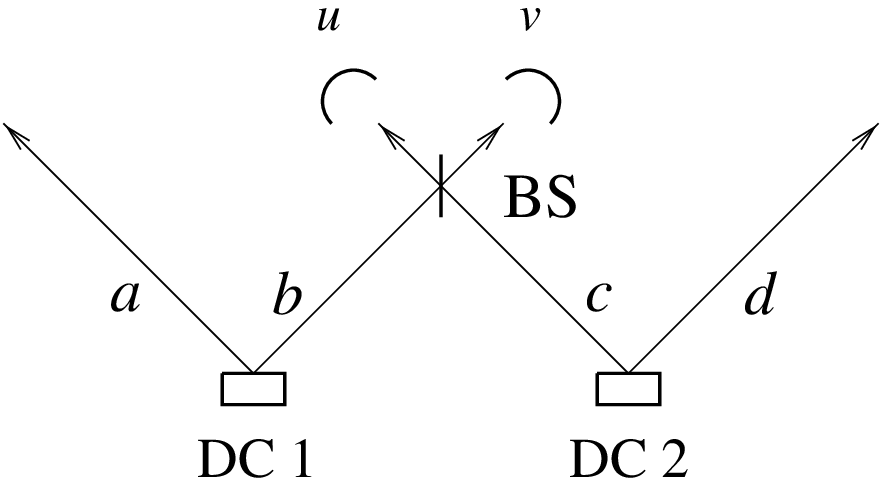}
  \end{psfrags}
 \end{center}
 {\small Fig.\ 1. A schematic representation of the experimental setup. Two 
	down-converters (DC) create states which exhibit polarisation 
	entanglement. One branch of each source is sent into a beam-splitter 
	(BS) and a coincidence in detectors $D_u$ and $D_v$ ideally identify 
	the $|\Psi^-\rangle$ Bell state.}
\end{figure}

\end{multicols}

\end{document}